\documentclass[superscriptaddress,twocolumn,showpacs,preprintnumbers,amsmath,amssymb]{revtex4}
\usepackage{graphicx}
\usepackage{dcolumn}
\usepackage{bm}
\usepackage{color}
\usepackage{ulem}
\usepackage{epsfig}

\begin{document}
\title{Intracavity frequency-doubled degenerate laser }

\author{Seng Fatt Liew}
\author{Sebastian Knitter}
\affiliation{Department of Applied Physics, Yale University, New Haven, CT 06520}
\author{Sascha Weiler}
\author{Jesus Fernando Monjardin-Lopez}
\author{Mark Ramme}
\affiliation{TRUMPF Inc, Farmington, CT 06032}
\author{Brandon Redding}
\affiliation{Department of Applied Physics, Yale University, New Haven, CT 06520}
\author{Michael A. Choma}
\affiliation{Deparment of Radiology and Biomedical Imaging, Yale University, New Haven, CT 06520}
\affiliation{Department of Pediatrics, Yale University, New Haven, CT 06520}
\affiliation{Department of Biomedical Engineering, Yale University, New Haven, CT 06520}
\affiliation{Department of Applied Physics, Yale University, New Haven, CT 06520}
\author{Hui Cao}
\affiliation{Department of Applied Physics, Yale University, New Haven, CT 06520}
\affiliation{Corresponding author: hui.cao@yale.edu}



\begin{abstract}
We develop a green light source with low spatial coherence via intracavity frequency doubling of a solid-state degenerate laser. 
The second harmonic emission supports many more transverse modes than the fundamental emission, and exhibit lower spatial coherence.
A strong suppression of speckle formation is demonstrated for both fundamental and second harmonic beams. 
Using the green emission for fluorescence excitation, we show the coherent artifacts are removed from the full-field fluorescence images. 
The high power, low spatial coherence and good directionality makes the green degenerate laser an attractive illumination source for parallel imaging and projection display. 
\end{abstract}


\maketitle


\indent Visible lasers have found a wide range of applications in imaging, spectroscopy and display. 
Their brightness, power efficiency and emission directionality make them more suitable than light emitting diodes (LEDs) for micro-projector applications in mobile devices. 
Unfortunately, the high coherence of a conventional laser produces speckle noise due to interference of the scattered light from a projection screen. 
For fluorescence microscopy, visible lasers are often used as excitation sources, because the common fluorescent dyes have absorption bands in the visible spectrum. 
Even though the fluorescence is spatially incoherent and does not produce speckle, coherent artifacts produced by the pump laser result in spatially non-uniform illumination which then corrupts the fluorescence image.
Coherent artifacts can be introduced by aberration of the imaging optics, and interface reflection or edge diffraction. 
To remove the coherent artifacts, various compounding techniques have been developed, e.g., using diffractive  elements \cite{Wang_AO98}, or rotating/vibrating components \cite{Yurlov_AO08,Kubota_AO10,Furue_JAP11}. 
Since these methods involve mechanically moving parts, they are relatively slow and require long acquisition times, which limit the application for high-speed imaging or projection. 

\indent A different approach to prevent speckle formation is developing lasers with low spatial coherence. 
A careful design of the laser cavity can facilitate lasing in many spatial modes with distinct emission pattern. 
The total emission from those mutually incoherent lasing modes has low spatial coherence, thus enabling speckle-free full-field imaging. 
To date, several types of such lasers have been developed, including dye-based random lasers \cite{Redding_NP12}, powder-based random Raman lasers \cite{Hokr_NC14}, solid-state degenerate lasers \cite{Nixon_OL13}, semiconductor-based chaotic microcavity lasers \cite{Redding_PNAS15}, broad-area vertical cavity surface emitting lasers (VCSELs) \cite{Riechert_OC08,Craggs_IEEE09,Verschaffelt_IEEE09}, VCSEL arrays \cite{Redding_OL14}, and degenerate vertical external cavity surface emitting lasers (VECSELs)  \cite{Knitter_Optica16}. 
All of these lasers except the dye-based random lasers \cite{Redding_NP12} and the powder-based random Raman lasers\cite{Hokr_NC14} have emission beyond the visible spectrum, making them unsuitable for imaging or display applications that require visible light. 
The dye-based random lasers and the powder-based random Raman lasers, though emitting in the visible spectrum, have high lasing threshold and poor collection efficiency. 
An alternative way of making visible sources, especially of green color, is frequency doubling of infrared (IR) lasers. 
Multi-wavelength, broadband green light sources have been developed for speckle noise reduction by spectral compounding \cite{Kuksenkov_SPIE11, Yu_OE14}, but there is a trade-off between the spectral bandwidth and the conversion efficiency, limiting the overall conversion efficiency.

Here, we propose frequency-doubling of a highly multimode IR laser to generate green emission with low spatial coherence. 
However, it is challenging to achieve an efficient conversion. 
Compared to a single-mode IR laser, the multimode laser emission has power distributed over many spatial modes, thus the power per mode is relatively low. 
Because of low spatial coherence, the laser beam cannot be focused tightly into a nonlinear crystal. 
Hence, the second harmonic generation (SHG) outside the laser cavity is expected to be inefficient, especially for the continuous-wave (CW) laser. 

To overcome these issues, we place a nonlinear crystal inside a degenerate laser cavity for frequency doubling. 
The high power density inside the laser cavity greatly enhances the SHG efficiency.
Previously, frequency doubling within a degenerate cavity was used to enhance the efficiency of second harmonic generation of weak images \cite{Treps_OE10}.
In this paper, we experiment with a CW solid-state disk laser that support more than 1000 transverse modes lasing simultaneously. 
Since lasing in different modes is not phase-locked, each mode is expected to produce its own second harmonic, thus the number of transverse modes in the second harmonic emission (at 532 nm) should be limited by the number of modes in the fundamental emission (at 1064 nm). 
Experimentally we find the second harmonic beam has many more transverse modes ($\sim$ 3300) than the fundamental beam ($\sim$ 1500). 
After transmitting through a static diffuser, the speckle contrast for the green light is notably lower than that of the IR. 
A physical explanation is provided for this surprising result. 
As an initial demonstration, we use the degenerate green laser as excitation source for fluorescence imaging. 
The low spatial coherence results in a significant improvement of the image quality as compared to that taken with a single-mode green laser. 
 
\begin{figure}[h]
	\centering
	\includegraphics[width=\linewidth]{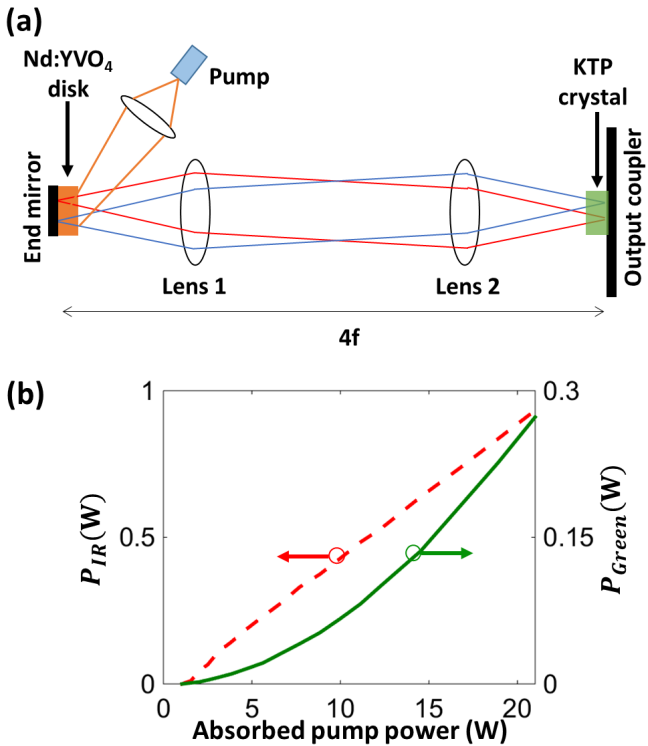}
	\caption{(Color online) (a) Schematic of the degenerate laser cavity with Nd:YVO$_4$ thin disk as the gain medium, two identical bi-convex lenses arranged in 4-f configuration, a KTP crystal for second harmonic generation, and an output coupler. The pump beam from a fiber coupled diode laser is focused to the Nd:YVO$_4$ crystal by a lens.  (b) Output power of fundamental emission at 1064 nm (red dashed line) and of the second harmonic emission at 532 nm (green solid line) vs. the absorbed pump power. The lasing threshold is 1.5 W. Above the threshold, the fundamental emission increases linearly with the pump and the second harmonic grows quadratically.}
	\label{fig1}
\end{figure}

\indent Figure \ref{fig1}(a) is a schematic of our Nd-vanadate degenerate laser with intracavity frequency doubling. 
The gain medium is a thin disk of $1 \%$ Nd:YVO$_4$ crystal (0.6 mm thick and 20 mm in diameter). 
One surface of the disk has high reflectivity (HR) coating to serve as an end mirror of the cavity. 
The other end mirror acts as an output coupler. 
In between them, two identical fused-silica plano-convex lenses with focal length $f$ = 100 mm are arranged in a 4$f$ telescope configuration. 
Each point on the disk surface will be imaged to the output coupler. 
In this self-imaging cavity, any transverse field distribution will be imaged onto itself after a single round trip. 
Therefore, the eigenmodes of the cavity can have arbitrary transverse field distribution and degenerate frequency. 
Since all the modes have the same quality factors, lasing can occur simultaneously in many transverse modes without phase correlation between them. 

A fiber coupled diode laser at 808 nm (Coherent, FAP800-40W) is used to pump  the Nd-vanadate crystal to the highly absorbing $^4F_{5/2}$ level. 
The pump beam is focused by a lens of $f$ = 30 mm to a circular spot of diameter $\sim 2$ mm on the front surface of the Nd-vanadate disk. 
The back surface of the disk has HR coating not only at the emission wavelength (1064 nm) but also at the pumping wavelength (808 nm). 
The incident pump beam is reflected from the back side and passes the disk twice, resulting in $70\%$ absorption of the incident pump light.

The Nd-vanadate disk is in contact with a heat sink through its HR coated surface. 
The heat sink is cooled with water at $20^{\circ}$ C. 
The thin disk geometry of the laser crystal greatly reduces the thermal lensing, because the heat flow direction is parallel to the laser cavity axis. 
The temperature in the radial direction of the disk is nearly uniform within the pump volume, only in the axial direction is a temperature gradient present \cite{Pavel_AO07}. 
The efficient cooling enables high-power CW lasing. 
Using an output coupler of $95\%$ reflectivity at 1064 nm, we obtain an output power of 7 W at 1064 nm when the (absorbed) pump power is 21 W. 

For intracavity frequency doubling, we choose the KTP crystal (KTiPO$_4$), which has a large angular bandwidth for type II phase matching \cite{Driscoll_JOSAB86}. 
Moreover, the KTP crystal has high thermal conductivity and low thermally induced aberration loss \cite{Perkins_JOSAB87}. 
The crystal is placed in close proximity to the output coupler. 
To enhance the intracavity power density at the fundamental frequency, we switch to an output coupler with $>99\%$ reflectivity at 1064 nm. 
For an efficient coupling of the second harmonic light out of the cavity, the output coupler has high transmissivity $\sim 98\%$ at 532 nm. 

Figure \ref{fig1}(b) plots the output powers at fundamental and second harmonic frequencies as a function of the (absorbed) pump power. 
The KTP crystal used for this experiment is 5 mm thick and has a lateral dimension of $9\times 9$ mm. 
Both sides of the crystal have anti-reflection (AR) coating at 1064 nm and 532 nm. 
To mitigate heating effect in the laser cavity, the CW pump beam is chopped to $10\%$ duty cycle with 60 Hz repetition rate. 
Above the lasing threshold of 1.5 W, the output power at the fundamental frequency grows linearly with the pump power, while the second harmonic displays a quadratic increase. 
This result indicates the linear loss of the degenerate laser cavity is dominant over the nonlinear loss due to SHG \cite{Smith_SHG_theory}. 
The second harmonic emission is linearly polarized with an extinction ratio of 50:1. 

\begin{figure}[htbp]
	\centering
	\includegraphics[width=\linewidth]{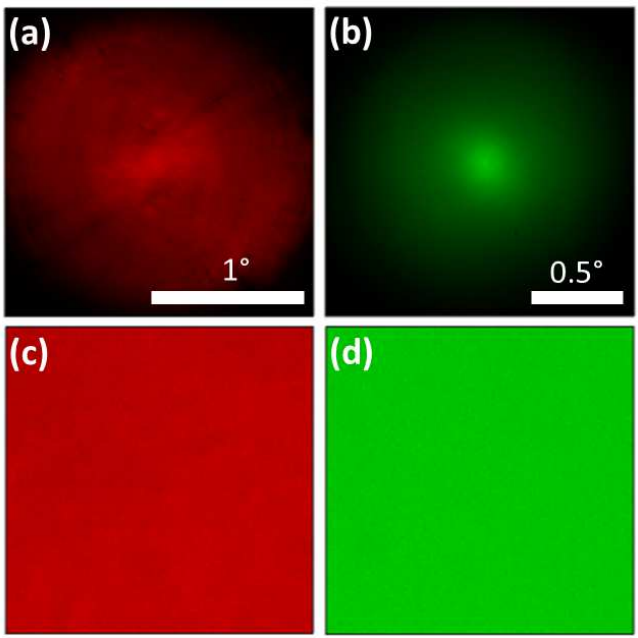}
	\caption{(Color online) (a,b) Far-field pattern (angular spectrum) of the fundamental (a) and second harmonic (b) output beams. (c,d) Speckle patterns produced by the fundamental (c) and second harmonic (d) beams after transmitting through a polarization maintaining ground-glass diffuser. The speckle contrast is 0.025 in (c) and 0.017 in  (d).}
	\label{fig2}
\end{figure}

\indent To determine the number of lasing modes, we measure the output beam quality at 1064 nm and 532 nm. 
The beam quality factor $M^2$ is given by the product of the near-field beam width and the far-field divergence, normalized to that of a single (transverse) mode beam \cite{Siegman_SPIE93}. 
We image the laser emission profile at the output coupler to compute the near-field beam width.
To measure the beam divergence, we put a lens at one focal length away from the output coupler, and record the far-field pattern (angular spectrum) at the back focal plane. 
The angular spectrum for the fundamental beam has maximum intensity at the center and a slowly decaying tail, as shown in Fig. \ref{fig2}(a), and the divergence angle is about 1$^\circ$.  
The angular spectrum for the second harmonic beam, as shown in Fig. \ref{fig2}(b), has a more pronounced peak at the center compared to the fundamental.
This result can be understood from the relation between the far field of second harmonic $E_{2 \omega}$ and that of fundamental $E_{\omega}$, i.e., $E_{2 \omega} = E_{\omega}\ast E_{\omega}$. 
The convolution relationship creates a sharper peak in the far field of the second harmonic. 


\indent By inserting a pinhole (of diameter 100 $\mu$m) to the degenerate cavity and placing it at the focal plane in between the two lenses, we increase the loss for high-order transverse modes and achieve lasing only in the fundamental mode \cite{Nixon_OL13}. 
To compute its space-beamwidth product, both near-field and far-field patterns of the single-mode laser are measured. 
By normalizing the space-beamwidth product of the multimode laser (without a pinhole in the cavity) to that of the single-mode laser, we obtain the beam quality factor, which gives the number of transverse modes at 1064 nm $N_{\omega}$ and 532 nm $N_{2 \omega}$. 
At the pump power of 21 W, $N_{\omega} =1500$ and $N_{2 \omega} = 3300$. 

\indent The presence of a large number of transverse modes in the laser emission reduces the spatial coherence, and prevent the speckle formation.
To confirm the speckle suppression, we direct the laser beam to a polarization maintaining ground-glass diffuser and record the speckle patterns of the transmitted light at 1064 nm and 532 nm separately, as shown in Fig. \ref{fig2}(c,d). 
The speckle contrast is calculated, $C = \sigma/\langle I\rangle$, where $\sigma$ is the standard deviation and $\langle I\rangle$ is the mean of all pixel intensities in the image. 
The speckle pattern at 1064 nm has low contrast $C$ = 0.025, but that at 532 nm has even lower contrast $C$ = 0.017. 
From the speckle contrast $C$, we estimate the number of spatial modes $N = 1/C^2$ to be 1600 in the fundamental beam and 3460 modes in the second harmonic beam. 
These values are close to those obtained from the beam quality factor measurement described above. 
The minor discrepancy is attributed to the error in the measurement of low speckle contrast due to the measurement noise, e.g., the camera noise. 
{Moreover, the finite polarization extinction ratio of 50:1 also reduces the speckle contrast slightly and gives higher estimation for the number of transverse modes than for perfectly polarized light. 

\begin{figure}[htbp]
	\centering
	\includegraphics[width=\linewidth]{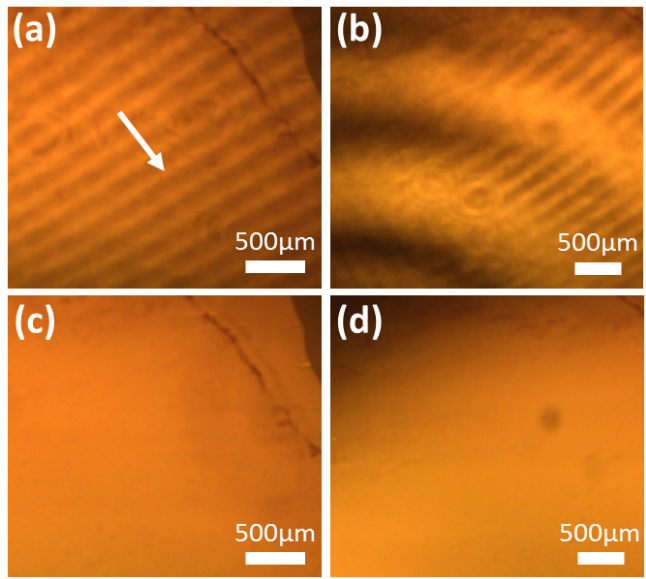}
	\caption{(Color online) Fluorescence images of dye-filled thin films excited by a single transverse mode green laser (a,b) or the frequency-doubled degenerate laser with low spatial coherence (c,d). In both cases, the excitation area on the sample surface is about 1 cm, and the incident pump intensity is around 1 mW/cm$^2$. (a) and (c) are taken of the same sample location, (b) and (d) of another location. The parallel fringes in (a) (marked by arrow) and rings in (b) result from the spatial modulations of pump intensity, which are removed in (c,d) when the spatial coherence of the pump laser is greatly reduced.  }
	\label{fig3}
\end{figure}  

\indent Since individual transverse modes in the degenerate cavity lase independently, they produce their own second harmonic. 
SHG is a coherent process, and second harmonic photons should not be generated from fundamental photons in mutually incoherent modes. 
As a result, one would expect the number of second harmonic modes is equal to or lower than that of the fundamental, but both beam quality factor and speckle contrast measurements show that the second harmonic has many more modes than the fundamental. 
In the degenerate cavity, the number of transverse modes at the second harmonic frequency is four times of that at the fundamental frequency, because the mode area of the second harmonic is four times smaller.  
The laser emission has a finite coherence time, which exceeds the round-trip time of the cavity. 
Different transverse modes are coherent within the coherence time, which is much longer than the passage time through the nonlinear crystal. 
Hence, the fundamental fields in different transverse modes are phase coherent while propagating through the KTP crystal, and they can generate the second harmonic fields.   
Due to large acceptance angle of the KTP crystal, two fundamental modes with slightly different transverse spatial frequencies $k_1 , k_2$ still satisfy the phase matching condition for SHG, and they generate the second harmonic field with the transverse spatial frequency of $(k_1+k_2)$, in addition to their own second harmonic fields of $2k_1$ and $2k_2$. 
Thus the second harmonic contains more transverse modes than the fundamental, but the ratio of their mode number is less than four because the higher-order transverse modes have lower nonlinear conversion efficiency than the lower order ones due to the angular dependent phase matching condition. 
    
\indent Finally, we use the green emission from the degenerate laser to remove coherent artifact in fluorescence imaging. 
For an initial demonstration, we make a fluorescence screen from 5 mMol Rhodamine 640 in diethylene glycol (DEG). 
The dye solution is injected to a 50-$\mu m$-thick chamber sandwiched between a cover slip and a glass slide. 
The four sides of the chamber are sealed with vacuum gel to prevent evaporation.}
The second harmonic output of the degenerate laser was focused to a spot of 10 mm on the screen to excite the dye molecules. 
The fluorescence image is recorded by a color camera (Moticam 2300). 
For comparison, we also use a commercial DPSS (diode-pumped solid-state) laser with single transverse mode as the excitation source. 
Figures \ref{fig3}(a) and (b) are fluorescence images taken with the single-mode laser illumination. 
Interference fringes and rings are clearly seen in the images. 
They originate from the spatial variations of the pump intensity. 
Such variations are caused by interference of multiple reflections from the imaging optics and the sample substrate of the coherent laser beam.
These coherent artifacts, however, vanish when the sample is illuminated by the second harmonic emission of the degenerate laser, as seen in Fig. \ref{fig3}(c, d). 
The low spatial coherence eliminates the interference-induced modulation of the excitation intensity.  

\indent In summary, we develop a green light source with low spatial coherence through intracavity frequency doubling of a thin-disk solid-state degenerate laser. 
The second harmonic emission contains $\sim$ 3300 transverse modes, and exhibits lower spatial coherence than the fundamental emission.  
Such low coherence reduces the speckle contrast to below $2 \%$, and removes coherent artifacts in the fluorescence imaging. 
The laser output power can be further increased by better matching the nonlinear loss to the linear loss of the degenerate cavity \cite{Smith_SHG_theory}. 
In the current system, half of the second harmonic emission, produced by the fundamental beam that is reflected from the output coupler, is lost. 
An HR coating on the distal surface of the nonlinear crystal at the second harmonic frequency will reflect such second harmonic emission to the output coupler and double the output power. 
Finally, the thermal load can be reduced by shifting the pump wavelength to 880 nm to pump directly to the $^4F_{3/2}$ emitting level. 
Although absorption of pump in a single pass is reduced, recirculating the pump beam in the laser crystal in a multipass pumping scheme will enhance the absorption efficiency \cite{Pavel_AO07}. 
The high power, low spatial coherence and good directionality will make the green degenerate laser an attractive illumination source for full-field imaging and projection display. 

\section*{Funding Information}
This work is supported by the National Institutes of Health (NIH) under the Grant No. 1R21HL125125-01A1, and by the MURI grant No. N00014-13-1-0649 from the US Office of Naval Research (ONR). 


\begin{thebibliography}{99}
\bibitem{Wang_AO98}
L. Wang, T. Tschudi, T. Halld\'{o}rsson, and P. R. P\'{e}tursson, ``Speckle reduction in laser projection systems by diffractive optical elements,'' Appl. Opt. \textbf{37}, 1770--1775 (1998).
\bibitem{Yurlov_AO08}
V. Yurlov, A. Lapchuk, S. Yun, J. Song, and H. Yang, ``Speckle suppression in scanning laser display,'' Appl. Opt. \textbf{47}, 179--187 (2008).
\bibitem{Kubota_AO10}
S. Kubota and J. W. Goodman, ``Very efficient speckle contrast reduction realized by moving diffuser device,'' Appl. Opt. \textbf{49}, 4385--4391 (2010).
\bibitem{Furue_JAP11}
H. Furue, A. Terashima, M. Shirao, Y. Koizumi, and M. Ono, ``Control of laser speckle noise using liquid
crystals,'' Jpn. J. Appl. Phys. \textbf{50}, 09NE14 (2011).
\bibitem{Redding_NP12}
B. Redding, M. A. Choma, and H. Cao, ``Speckle-free laser imaging using random laser illumination,'' Nat. Photonics \textbf{6}, 355 (2012). 
\bibitem{Hokr_NC14}
B. H. Hokr, J. N. Bixler, M. Cone, J. D. Mason, H. T. Beier, G. D. Noojin, G. I. Petrov, L. A. Golovan, R. J. Thomas, B. A. Rockwell, and V. V. Yakovlev, ``Bright emission from a random Raman laser'', Nat. Commun. \textbf{5}, 4356 (2014).
\bibitem{Nixon_OL13}
M. Nixon, B. Redding, A. A. Friesem, H. Cao, and N. Davidson, ``Efficient method for controlling the spatial coherence of a laser,'' Opt. Lett. \textbf{38}, 3858--3861 (2013).
\bibitem{Redding_PNAS15}
B. Redding, A. Cerjan, X. Huang, M. L. Lee, A. D. Stone, M. A. Choma, and H. Cao, ``Low spatial coherence electrically pumped semiconductor laser for speckle-free full-field imaging,'' Proc. Natl. Acad. Sci. USA \textbf{112}, 1304 (2015).
\bibitem{Riechert_OC08}
F. Riechert, G. Verschaffelt, M. Peeters, G. Bastian, U. Lemmer, and I. Fischer, ``Speckle characteristics of a broad-area VCSEL in the incoherent emission regime,'' Opt. Commun. \textbf{281}, 4424 (2008). 
\bibitem{Craggs_IEEE09}
G. Craggs, G. Verschaffelt, S. K. Mandre, H. Thienpont, and I. Fischer, ``Thermally Controlled Onset of Spatially Incoherent Emission in a Broad-Area Vertical-Cavity Surface-Emitting Laser,'' IEEE J. Sel. Top. Quantum Electron. \textbf{15}, 555 (2009). 
\bibitem{Verschaffelt_IEEE09}
G. Verschaffelt, G. Craggs, M. L. Peeters, S. K. Mandre, H. Thienpont, and I. Fischer, ``Spatially Resolved Characterization of the Coherence Area in the Incoherent Emission Regime of a Broad-Area Vertical-Cavity Surface-Emitting Laser,'' IEEE J. Quantum Electron. \textbf{45}, 249 (2009). 
\bibitem{Redding_OL14}
B. Redding, Y. Bromberg, M. A. Choma, and H. Cao, ``Full-field interferometric confocal microscopy using a VCSEL array,'' Opt. Lett. 39, 4446---4449 (2014)
\bibitem{Knitter_Optica16}
S. Knitter, C. Liu, B. Redding, M. K. Khokha, M. A. Choma, and H. Cao, ``Coherence switching of a degenerate VECSEL for multimodality imaging,'' Optica \textbf{3}, 403--406 (2016).
\bibitem{Kuksenkov_SPIE11}
D. V. Kuksenkov, R. V. Roussev, S. Li, W. A. Wood, and C. M. Lynn, ``Multiple-wavelength synthetic green
laser source for speckle reduction,'' Proc. SPIE \textbf{7917}, 79170B (2011).
\bibitem{Yu_OE14}
N. E. Yu, J. W. Choi, H. Kang, D.-K. Ko, S.-H. Fu, J.-W. Liou, A. H. Kung, H. J. Choi, B. J. Kim, M. Cha, and L.-H. Peng, ``Speckle noise reduction on a laser projection display via a broadband green light source,'' Opt. Express \textbf{22}, 3547--3556 (2014).
\bibitem{Treps_OE10}
B. Chalopin, A. Chiummo, C. Fabre, A. Ma\^{i}tre, and Nicolas Treps, ``Frequency doubling of low power images using a self-imaging cavity,'' Opt. Express \textbf{18}, 8033--8042 (2010).
\bibitem{Pavel_AO07}
N. Pavel, K. L\"{u}nstedt, K. Petermann, and G. Huber, ``Multipass pumped Nd-based thin-disk lasers: continuous-wave laser operation at 1.06 and 0.9 $\mu$m with intracavity frequency doubling,'' Appl. Opt. \textbf{46}, 8256--8263 (2007).
\bibitem{Driscoll_JOSAB86}
T. A. Driscoll, H. J. Hoffman, R. E. Stone, and P. E. Perkins, ``Efficient second-harmonic generation in KTP crystals,'' J. Opt. Soc. Am. B \textbf{3}, 683--686 (1986).
\bibitem{Perkins_JOSAB87}
P. E. Perkins and T. S. Fahlen, ``20-W average-power KTP intracavity-doubled Nd:YAG laser,'' J. Opt. Soc. Am. B \textbf{4}, 1066--1071 (1987).
\bibitem{Smith_SHG_theory}
R. G. Smith, ``Theory of intracavity optical second-harmonic generation,'' IEEE J. Quantum Electron. \textbf{QE-6}, 215 (1970). 
\bibitem{Siegman_SPIE93}
A. E. Siegman, ``Defining, measuring, and optimizing laser beam quality,'' Proc. SPIE \textbf{1868}, 2 (1993). 
\end{thebibliography}



\end{document}